# Gauge field and geometric control of quantum-thermodynamic engine

Sumiyoshi Abe[1,2]

[1] Department of Physical Engineering, Mie University, Mie 514-8507, Japan

[2] Institut Supérieur des Matériaux et Mécaniques Avancés, 44 F. A. Bartholdi, 72000 Le Mans, France

**Abstract.** The problem of extracting the work from a quantum-thermodynamic system driven by slowly varying external parameters is discussed. It is shown that there naturally emerges a gauge-theoretic structure. The field strength identically vanishes if the system is in an equilibrium state, i.e., the nonvanishing field strength implies that the system is in a nonequilibrium quasi-stationary state. The work done through a cyclic process in the parameter space is given in terms of the flux of the field. This general formalism is applied to an example of a single spin in a varying magnetic field, and the maximum power output is discussed in a given finite-time cyclic process.

**Keywords:** Non-equilibrium processes

# 1. Introduction

Developments in manipulating microscopic systems have been drawing fresh interest in quantum thermodynamics and thermodynamics of small systems [1]. This is indeed the case from nano scales to even scales of a few particles. There, extracting the work from a system by controlling external conditions is one of primary issues. In particular, it is of importance to consider such a problem in nonequilibrium situations, if a variety of examples ranging from condensed matter systems to (bio)molecular motors are taken into account (see, e.g., Ref. [2]). Thus, any physical idea that can suggest a new experimental method of controlling nonequilibrium small systems is currently welcome.

We would also like to mention the study in Ref. [3]. It is shown in that study how the work can be extracted from a single particle confined in a potential well by varying quantum states and the width of the well in a peculiar way. This is a pure-state quantum-mechanical analog of the Carnot engine, but no environments are employed. Recently, such a similarity between quantum mechanics and thermodynamics was further clarified in Re. [4]. In addition, the exact analytical expression was derived for the efficiency of the engine at its maximum power output [5].

In this paper, we discuss a possibility of *geometrically* controlling systems in the quantum-mechanical regime. We focus our attention on the gauge-theoretic nature of the work in both equilibrium and nonequilibrium quantum thermodynamics. There emerges a gauge field when a system is driven by *slowly varying* external parameters. The domain of this field is the parameter space. The field, in fact, transforms as an Abelian gauge field under a class of redefinitions of the system density matrix. The field is pure gauge, that is, its strength identically vanishes, if the system is in an equilibrium state. In other words, the nonvanishing field strength implies that the state is out of equilibrium. Thus, the nonequilibrium nature is characterized in a novel manner. Of particular interest is a cyclic process described by a closed curve in the parameter space. The thermodynamic work done along such a process is given in terms of the flux of the gauge field penetrating the surface surrounded by the closed curve, which is nonvanishing if and only if the system is in a nonequilibrium state. We apply this general scheme to an example of a single spin in a rotating magnetic field, and discuss the condition of the maximum power output in a finite-time process.

## 2. Work and gauge field

Consider a driven quantum system under an external condition. The system Hamiltonian has the following form:

$$H = H(\lambda^1, \lambda^2, ..., \lambda^n). \tag{1}$$

$\lambda$'s are the parameters representing the external driving and are regarded as a coordinate point in the *n*-dimensional parameter space, $M$. Let $\rho$ be a density matrix describing the state of the system, which is a normalized Hermitian positive semidefinite matrix. The internal energy reads $U = \text{Tr}(\rho H)$. Its change along a certain process is $dU = \text{Tr}(d\rho\, H) + \text{Tr}(\rho\, dH)$, which yields the first law of thermodynamics, $d'Q = dU + d'W$, if changes of the quantity of heat and work, $d'Q$ and $d'W$, are identified with $\mathrm{Tr}(d\rho\, H)$ and $-\mathrm{Tr}(\rho\, dH)$, respectively.

From Eq. (1), we have

$$d'W = a_i\, d\lambda^i, \qquad (2)$$

$$a_i = -\mathrm{Tr}(\rho\, \partial_i H), \qquad (3)$$

where $\partial_i \equiv \partial/\partial \lambda^i$ and Einstein's summation convention is understood for the repeated indices. In a thermodynamic situation, the change of $\lambda$'s in time should be very slow compared to the microscopic dynamical time scale, and $\rho$ is assumed to be a quasi-stationary state. In such a case, $a_i$'s become the components of a gauge field potential, as seen below.

The work done along a curve, $\Gamma$, in $M$ is given by $W_\Gamma = \int_\Gamma d\lambda^i\, a_i$. Here, let us consider a cyclic process along a closed curve $C$ in $M$. Then, using the Stokes theorem, we find the work to be

$$W_C = \oint_C d\lambda^i\, a_i = \frac{1}{2} \iint_S d\lambda^i \wedge d\lambda^j\, f_{ij}, \qquad (4)$$

where $S$ is a surface surrounded by $C$ (that is, $\partial S = C$), and $f_{ij}$ is the field strength defined by

$$f_{ij} = \partial_i a_j - \partial_j a_i. \qquad (5)$$

Eq. (4) holds, since the change of $\lambda$'s in time is sufficiently slow. Thus, the work is given by the flux penetrating $S$. It is clear that the net entropy change after completion of one cycle is zero.

Now, let us see that $a_i$ in Eq. (3) can be regarded as a gauge field. The density matrix may transform as follows:

$$\rho \rightarrow \rho + \tilde{\rho}, \tag{6}$$

where $\tilde{\rho}$ is traceless. $\tilde{\rho}$ should not violate the positive semidefiniteness of the transformed density matrix. Applying the transformation in Eq. (6) to Eq. (3), we find that $a_i$ transforms as an Abelian gauge field

$$a_i \rightarrow a_i + \partial_i \Lambda, \tag{7}$$

if $\tilde{\rho}$ is a function only of the Hamiltonian, i.e., $\tilde{\rho} = \tilde{\rho}(H)$, or is independent of $\lambda$'s. This leaves the field strength in Eq. (5) invariant. An example of $\tilde{\rho}(H)$ is $\tilde{\rho}(H) = \left(H - \langle H \rangle_0\right)\rho_0(H)$, where $\rho_0(H)$ is a unit-trace matrix depending only on $H$ and $\langle H \rangle_0 \equiv \text{Tr}(\rho_0 H)$. $\Lambda$ is the function satisfying $\partial_i \Lambda \equiv -\text{Tr}[\tilde{\rho}(H) \partial_i H]$.

We note however that the concept of gauge invariance is actually somewhat restricted. Consider the internal energy, for example. It transforms as

$$U = \text{Tr}(\rho H) \rightarrow \text{Tr}(\rho H) + \text{Tr}(\tilde{\rho} H) \equiv \text{Tr}(\rho H'), \tag{8}$$

where $H' = H + c$ with $c$ being $c = \text{Tr}(\tilde{\rho} H)$. Thus, the gauge invariance is realized up to the $c$-number shift of the Hamiltonian.

The above discussion characterizes nonequilibrium states in a peculiar manner. Suppose that $\rho$ in Eq. (3) is a function only of the Hamiltonian, then the field strength in Eq. (5) identically vanishes, because of the total derivative nature of $a_i$ (i.e., the pure gauge). For example, let us consider the canonical density matrix of an equilibrium state

$$\rho_{eq}(H) = \frac{1}{Z(\beta)} \exp(-\beta H) \tag{9}$$

with the partition function

$$Z(\beta) = \mathrm{Tr} \exp(-\beta H) \tag{10}$$

and $\beta$ being the inverse temperature. In this case, we find that $a_i$ is the pure gauge:

$$a_i = -\partial_i F, \tag{11}$$

where $F = -\beta^{-1} \ln Z(\beta)$ is the Helmholtz free energy. So, the field strength in Eq. (5), in fact, vanishes, and Eq. (2) yields $d'W = -dF$, implying a well-known fact that the work done is given by the free energy difference in an isothermal process. Therefore, if the system is in an equilibrium state, then the field strength identically vanishes. Contrarily, a nonvanishing field strength implies that the corresponding state of the system is out of equilibrium. This offers a gauge-theoretic characterization of the nonequilibrium nature.

In a particular case when a system is in a pure state, say, $\rho = |\psi\rangle\langle\psi|$, the gauge field is given by $a_i = -\langle\psi|\partial_i H|\psi\rangle$. Note that, in this case, Eq. (6) is a transformation from a pure state to a mixed state, in general.

## 3. Example: Finite-time thermodynamics of a spin in a rotating magnetic field

Now, we examine the formalism developed in the preceding section for a simple example of a spin-1/2 in a varying magnetic field, $\mathbf{B}$. Although it may actually be difficult to realize a quasi-stationary state of such a system, we examine it here just as a prototype example. The Hamiltonian reads

$$H = \hbar\kappa\,\sigma \cdot \mathbf{B}, \qquad (12)$$

where $\sigma$'s are the Pauli matrices and $\kappa$ is a constant including the magnetic moment. Henceforth, $\hbar\kappa$ is set equal to unity. $\mathbf{B} = (B_x, B_y, B_z)$ plays a role of $\lambda$'s in Eq. (1), and thus the parameter space, $M$, is 3-dimensional. The adiabatic approximation allows the equation, $H|\pm(\mathbf{B})\rangle = E_\pm|\pm(\mathbf{B})\rangle$, to hold. The eigenstates corresponding to the eigenvalues, $E_\pm = \pm B$, are $|+(\mathbf{B})\rangle = [2B(B+B_z)]^{-1/2}(B+B_z, B_x+iB_y)^\mathrm{T}$ and $|-(\mathbf{B})\rangle = [2B(B+B_z)]^{-1/2}(B_x-iB_y, -B-B_z)^\mathrm{T}$, respectively, where $B = |\mathbf{B}|$. Let us consider the following general stationary mixed states:

$$\begin{aligned}\rho = {}& p|+(\mathbf{B})\rangle\langle+(\mathbf{B})| + (1-p)|-(\mathbf{B})\rangle\langle-(\mathbf{B})| \\ & + \alpha|-(\mathbf{B})\rangle\langle+(\mathbf{B})| + \alpha^*|+(\mathbf{B})\rangle\langle-(\mathbf{B})|,\end{aligned} \qquad (13)$$

where the condition $p(1-p) \geq |\alpha|^2$ should be fulfilled in order for $\rho$ to be positive semidefinite. The last two terms on the right-hand side highlight the nonequilibrium nature of the state. $p$ and $\alpha$ may depend on $\mathbf{B}$, in general. But, here, they are taken to be constant and, in addition, $\alpha$ to be real, for the sake of simplicity. Then, the gauge field in Eq. (3) becomes

$$\mathbf{a} = -p \langle +(\mathbf{B})|\sigma|+(\mathbf{B})\rangle - (1-p)\langle -(\mathbf{B})|\sigma|-(\mathbf{B})\rangle$$
$$- 2\alpha \mathrm{Re}\langle +(\mathbf{B})|\sigma|-(\mathbf{B})\rangle. \qquad (14)$$

The matrix elements are given as follows: $\langle \pm(\mathbf{B})|\sigma|\pm(\mathbf{B})\rangle = \pm \mathbf{B}/B$, $\langle +(\mathbf{B})|\sigma_x|-(\mathbf{B})\rangle$
$= -(B+B_z)/(2B) + (B_x - iB_y)^2/[2B(B+B_z)]$, $\langle +(\mathbf{B})|\sigma_y|-(\mathbf{B})\rangle = i(B+B_z)/(2B)$
$+ i(B_x - iB_y)^2/[2B(B+B_z)]$, $\langle +(\mathbf{B})|\sigma_z|-(\mathbf{B})\rangle = (B_x - iB_y)/B$. Therefore,

$$\mathbf{a}(\mathbf{B}) = (1-2p)\mathbf{B}/B - \alpha \begin{pmatrix} (B_x^2 - B_y^2)/[B(B+B_z)] - (B+B_z)/B \\ 2B_xB_y/[B(B+B_z)] \\ 2B_x/B \end{pmatrix}. \qquad (15)$$

Note that the first term on the right-hand side, which comes from the diagonal elements, is a part of the pure gauge. The field strength is given by

$$\mathbf{b}(\mathbf{B}) = \nabla_B \times \mathbf{a}(\mathbf{B}) = 2\alpha \begin{pmatrix} 0 \\ 1/B \\ -B_y/[B(B+B_z)] \end{pmatrix}, \qquad (16)$$

where $\nabla_B \equiv \partial/\partial \mathbf{B}$. Let us consider a cyclic variation of the magnetic field along a curve, $C$, on the $B_x B_y$-plane. Using the general formula in Eq. (4), we have

$$W_C = -2\alpha \iint_S dB_x dB_y \frac{B_y}{B_x^2 + B_y^2}. \tag{17}$$

Let us consider $C_1$ depicted in Fig. 1. It is a circle with radius $R$, and the process is clockwise. The integral in Eq. (17) is immediately performed to yield

$$W_{C_1} = 2\alpha \pi R. \tag{18}$$

This is, in a sense, a trivial case.

For the discussion about finite-time thermodynamics, a nontrivial case can be realized by the process in Fig. 2. The closed curve, $C_2$, is a rectangle with the lengths of the sides, $L_1$ and $L_2$. The work is calculated to be

$$W_{C_2} = \alpha L_1 \ln\left[1 + \left(\frac{L_2}{L_1}\right)^2\right] + 2\alpha L_2 \tan^{-1}\left(\frac{L_1}{L_2}\right). \tag{19}$$

This work depends on two parameters, $L_1$ and $L_2$, and therefore there is a possibility of maximizing the power output

$$P = \frac{W}{\tau}. \tag{20}$$

in the finite-time process. Here, $\tau$ stands for the cycle time. Let $v(t)$ be the speed of the variation of $\mathbf{B}$ along $C_2$. That is,

$$\int_0^\tau dt\, v(t) = \bar{v}\tau = 2(L_1 + L_2), \tag{21}$$

where $\bar{v}$ denotes the average speed. Since the process should be slow enough, $\bar{v}/(L_1 + L_2)$ has to be much smaller than the microscopic dynamical frequency, $|E_\pm|/\hbar$. We rewrite Eq. (20) as follows:

$$P = \frac{\alpha \bar{v}}{2} f(r), \tag{22}$$

where

$$r = \frac{L_1}{L_2}, \tag{23}$$

$$f(r) = \frac{r\ln\left(1 + \frac{1}{r^2}\right) + 2\tan^{-1} r}{1 + r}. \tag{24}$$

Therefore, controlling the ratio, $r$, we can maximize the power output. The plot of $f(r)$ with respect to $r$ is presented in Fig. 3. The maximum is observed around $r = r^* \cong 0.7$. The corresponding value of the work is

$$W^* = \alpha \sqrt{\frac{A}{r^*}} \left[ r^* \ln\left(1 + \frac{1}{r^{*2}}\right) + 2\tan^{-1} r^* \right], \tag{25}$$

where $A$ is the area, $A = L_1 L_2$, constrained by the ratio $L_1/L_2 = r^*$.

## 4. Conclusion

We have developed a gauge-theoretic discussion about the extraction of the work from a nonequilibrium quantum-thermodynamic system driven by slowly varying external parameters. We have described the work done along a closed curve in the parameter space as the flux of the gauge field. The nonequilibrium nature of quantum-thermodynamic states is characterized in the language of gauge theory. We have applied this formalism to an example of a spin in a varying external magnetic field, and have discussed the condition of the maximum power output in a given finite-time cyclic process in the parameter space. In the present study, the states are assumed to be quasi-stationary, and the state change is realized only through slow variation of the external parameters. It is of interest to explicitly introduce dynamical evolution of the state through a master equation of a certain kind. Such an approach will put a basis for the effective treatment of a finite-time process considered here.

## Acknowledgments

This research was supported in part by a Grant-in-Aid for Scientific Research from the Japan Society for the Promotion of Science and the Project of Knowledge Innovation Program (PKIP) of Chinese Academy of Sciences, Grant No. KJCX2.YW.W10. The author would like to thank the Kavli Institute for Theoretical Physics China for providing him with the nice working atmosphere and hospitality.

# Figure Caption

FIG. 1    A circular process on the $B_x B_y$ − plane. All quantities have the dimension of the magnetic flux density.

FIG. 2    A rectangular process on the $B_x B_y$ − plane. All quantities have the dimension of the magnetic flux density.

FIG. 3    The plot of $f(r)$ with respect to $r$. All quantities are dimensionless.

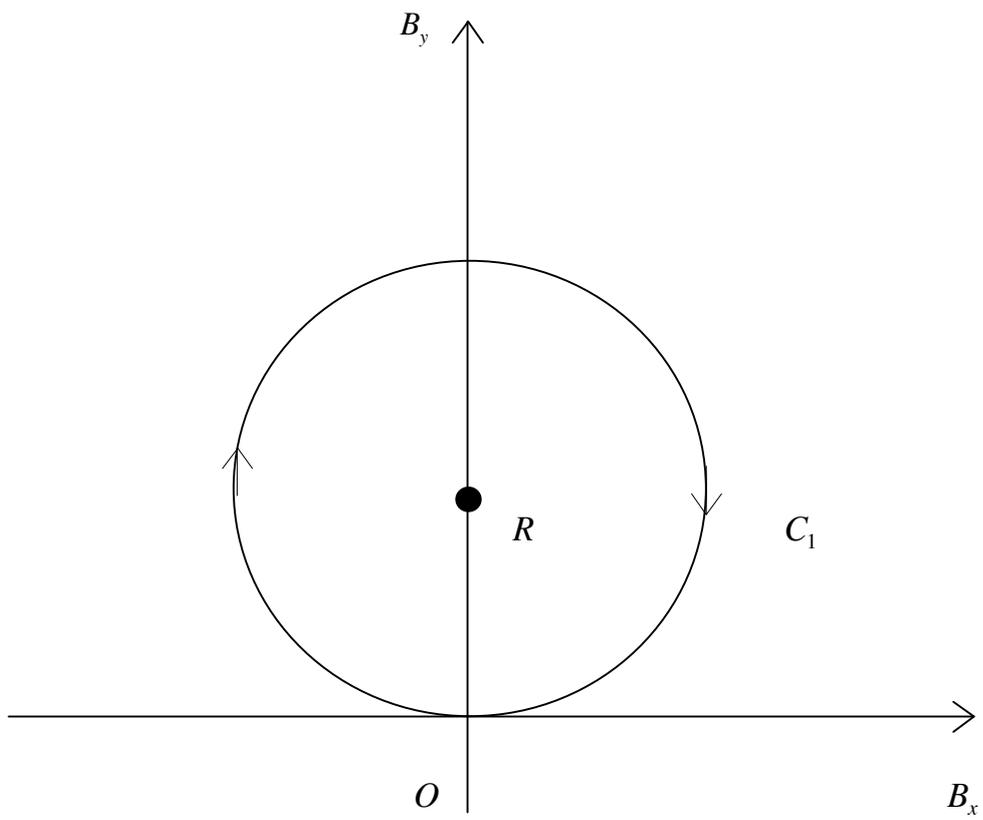

Figure 1

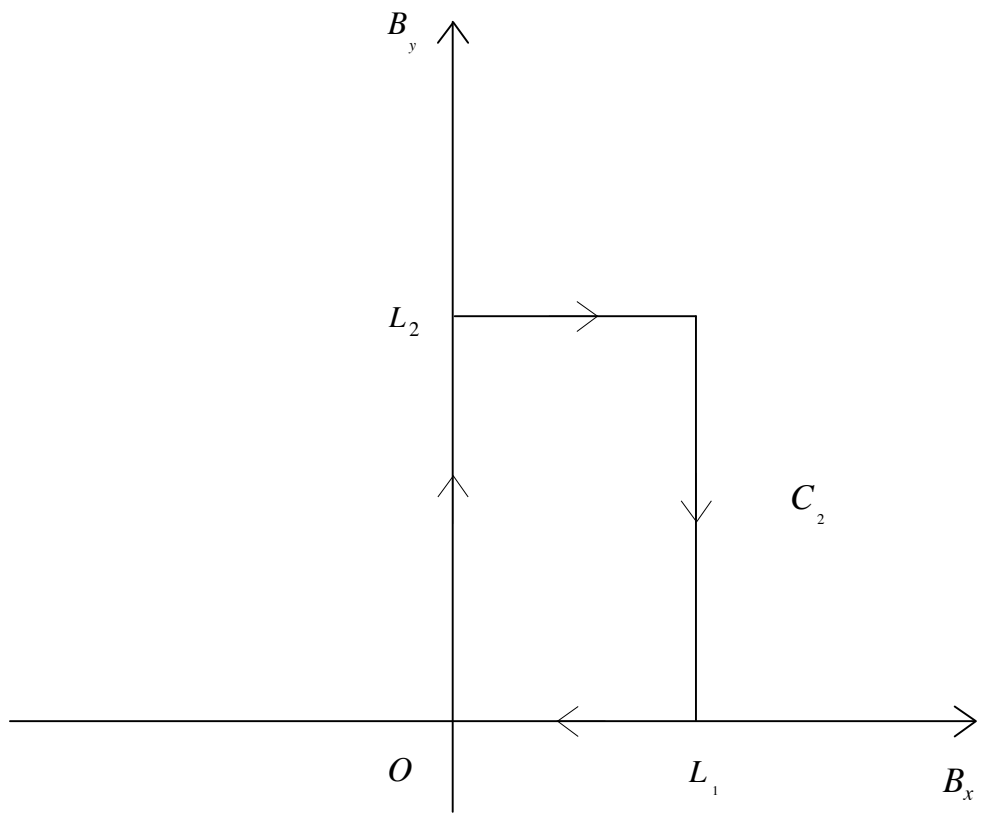

Figure 2

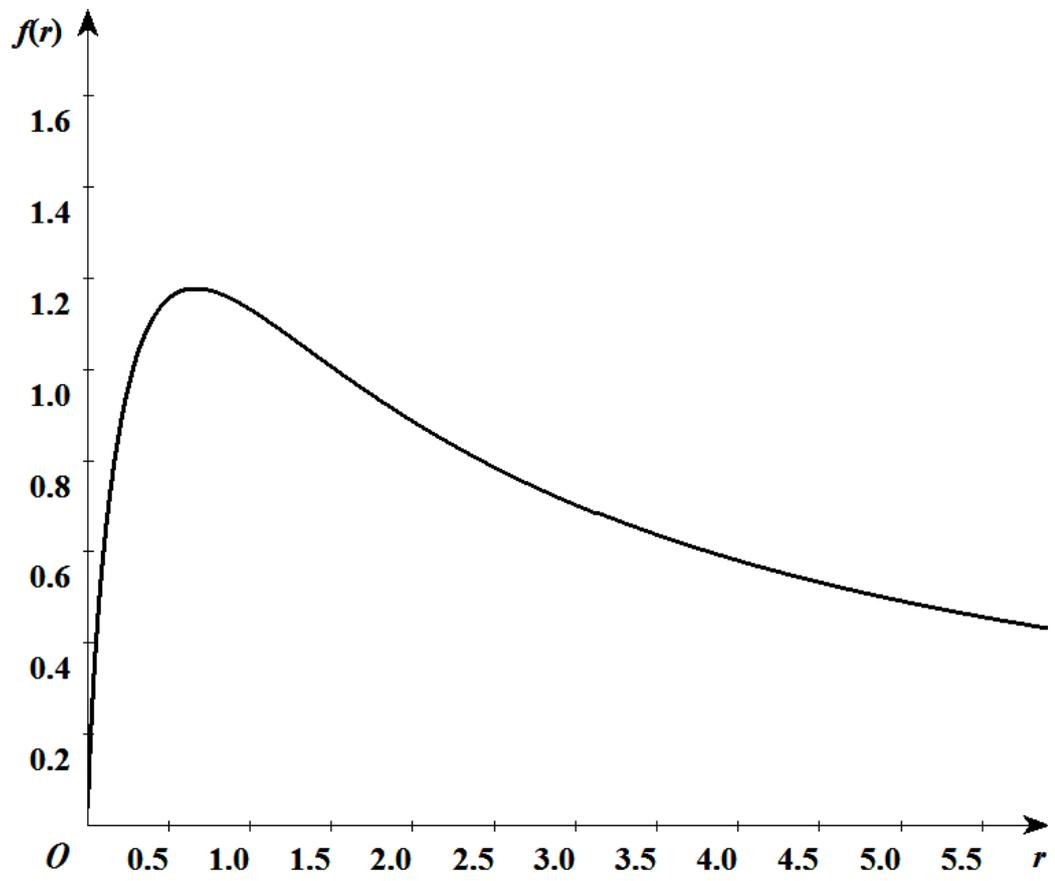

Figure 3